# Estimating Variability in Hospital Charges: The Case of Cesarean Section

Anna Perfilyeva[1], Vittal Raghavendra Miskin[1], Ryan Aven[1], Craig Drohan[1], and Huthaifa I. Ashqar[1,2,*]

[1] Department of Computer Science and Electrical Engineering, University of Maryland Baltimore County (UMBC)
[2] Arab American University, Jenin, Palestine
[*] Corresponding Author

**Abstract**
This study sought to better understand the causes of price disparity in cesarean sections, using newly released hospital data. Beginning January 1, 2021, Centers for Medicare and Medicaid Services (CMS) requires hospitals functioning in the United States to publish online pricing information for items and services these hospitals provide in a machine-readable format and a consumer friendly shoppable format.[1] Initial analyses of these data have shown that the price for a given procedure can differ in a hospital and across hospitals. The cesarean section (C-section) is one of the most common inpatient procedures performed across all hospitals in the United States as of 2018.[2] This preliminary study found that for C-section procedures, pricing varied from as little as $162 to as high as $115,483 for a single procedure. Overall, indicators for quality and whether or not the hospital was a teaching hospital were found to be significantly significant, while variables including median income and the gini coefficient for wealth inequality were not shown to be statistically significant.

---

[1] *Hospital Price Transparency*. CMS, December 2021, <https://www.cms.gov/hospital-price-transparency>. Last accessed May 5, 2022.
[2] *HCUP Fast Stats - Most Common Operations During Inpatient Stays*. Agency for Healthcare Research and Quality, April 2021, <https://www.hcup-us.ahrq.gov/faststats/NationalProceduresServle>. Last accessed April 29, 2022.

**Introduction: CMS, ACA, and the Long March to Price Transparency**
Historically, the prices hospitals charge for medical services and procedures have been frustratingly unknown for consumers until the moment they receive the final bill.[3] This is because hospitals typically do not publish their list of prices on a menu board as you might expect to see when you go into a fast-food establishment. Instead, the typical hospital pricing strategy has been opaque and convoluted, perhaps intentionally so. With limited pricing information publicly available hospitals did not have to compete on price. With increased price transparency, market competition could lower patient costs, as consumers could more easily compare prices by hospital and choose a lower-cost provider.[4] In other words, price transparency data could allow individuals to make better informed healthcare decisions and potentially lower their medical costs.[5]

Having access to this type of pricing transparency data is important for a number of reasons, mainly to do with helping consumers lower the financial burden of medical procedures that can lead to debt. A 2017 study from the Census Bureau estimated that 19 percent of US households carry medical debt with a median amount owed of $2,000.[6] A more recent study has estimated (after taking the medical costs of COVID-related issues into account) that the US as a whole has over $140 billion in total medical debt.[7] Health care spending in the US increased by 9.7 percent in 2020 (the most recent year available), compared to only 4.3 percent growth in 2019, significantly faster than workers' wages, which rose an average of 2.6 and 2.8 percent respectively.[8] This is especially concerning, as healthcare expenditures are already quite high in the United States, making up 20 percent of total spending in the US economy.[9] As shown in Figure 1 below,[10] healthcare expenditures per capita has steadily increased over the past 20 years and has in fact doubled from 2004 to 2020:

---

[3] Emanuel, Ezekiel, et al. "A Systemic Approach to Containing Health Care Spending." New England Journal of Medicine, vol. 367, no. 10, 2012, pp. 949–54. <https://www.nejm.org/doi/full/10.1056/NEJMsb1205901>.

[4] Ibid

[5] How New Data On Hospital "Discounted Cash Prices" Might Lead To Patient Savings. (2021). *Forefront Group*. <https://www.healthaffairs.org/do/10.1377/forefront.20211103.716124>.

[6] U.S. Census Bureau. (2021, December 16). *19% of U.S. Households Could Not Afford to Pay for Medical Care Right Away*. Census.Gov. <https://www.census.gov/library/stories/2021/04/who-had-medical-debt-in-united-states.html>.

[7] Kluender R, Mahoney N, Wong F, Yin W. Medical Debt in the US, 2009-2020. *JAMA*. 2021;326(3):250–256. <https://jamanetwork.com/journals/jama/fullarticle/2782187>.

[8] *National Expenditures Highlights*, CMS, 2020, <https://www.cms.gov/files/document/highlights.pdf>. Last accessed April, 28, 2022.

[9] Ibid

[10] *National Health Expenditure Accounts*, CMS, 2020. <https://www.cms.gov/Research-Statistics-Data-and-Systems/Statistics-Trends-and-Reports/NationalHealthExpendData/NationalHealthAccountsHistorical>. Last accessed May, 9, 2022.

**Figure 1: Healthcare Consumption Expenditures in the United States**

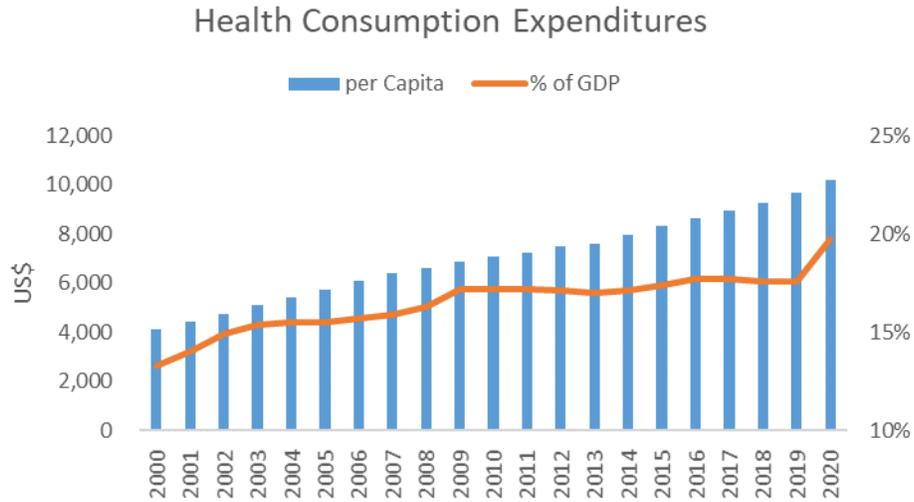

Source: CMS NHEA

In 2012, an eleven-point plan was released by a collection of the nation's top healthcare experts, which stated that the "full transparency of prices" was a critical piece of the solution to ameliorate steadily increasing healthcare costs.[11] The passage of the Affordable Care Act (ACA) and subsequent consolidation of health information enabled further movement towards transparent healthcare pricing. In 2018, the Centers for Medicare & Medicaid Services (CMS) required hospitals to post their chargemasters online in an attempt to provide patients better access to hospital price information.[12] A significant problem with this plan, however, is that chargemaster files do not necessarily reflect true costs of healthcare but rather are primarily used within hospitals' finance departments to help manage revenue cycles.[13] After recognizing this issue, in November 2019 CMS published "Price Transparency Requirements for Hospitals to Make Standard Charges Public," which required hospitals to post actual prices paid for common procedures instead of just the fee-for-service prices listed in the chargemaster.[14]

---

[11] Emanuel, Ezekiel, et al. "A Systemic Approach to Containing Health Care Spending." New England Journal of Medicine, vol. 367, no. 10, 2012, pp. 949–54. <https://www.nejm.org/doi/full/10.1056/NEJMsb1205901>.

[12] *Fiscal Year (FY) 2019 Medicare Hospital Inpatient Prospective Payment System (IPPS) and Long-Term Acute Care Hospital (LTCH) Prospective Payment System Final Rule (CMS-1694-F)*, CMS, August 2, 2018, <https://www.cms.gov/newsroom/fact-sheets/fiscal-year-fy-2019-medicare-hospital-inpatient-prospective-payment-system-ipps-and-long-term-acute-0>. Last accessed May 1, 2022.

[13] George Washington University's Online Healthcare MBA. (2020, November 5). *What Is a Chargemaster and What Do Hospital Administrators Need to Know About It?* George Washington University. <https://healthcaremba.gwu.edu/blog/chargemaster-hospital-administrators-need-know/>

[14] "New Year, New CMS Price Transparency Rule For Hospitals", Health Affairs Blog, January 19, 2021. <https://www.healthaffairs.org/do/10.1377/forefront.20210112.545531/full/>

In order to be compliant with the CMS mandate, hospitals must publish the following items in a machine-readable format:[15]
1. The gross charge: the charge for an individual item or service that is reflected on a hospital's chargemaster, absent any discounts
2. The discounted cash price: the charge that applies to an individual who pays cash, or cash equivalent, for a hospital item or service
3. The payer-specific negotiated charge: the charge that a hospital has negotiated with a third party payer for an item or service
4. The de-identified minimum negotiated charge: the lowest charge that a hospital has negotiated with all third-party payers for an item or service
5. The de-identified maximum negotiated charge: the highest charge that a hospital has negotiated with all third-party payers for an item or service

The first data available from the new CMS requirements began to be available for analysis in 2021, offering researchers and academics a growing trove of previously unavailable pricing information. In our study, we chose to analyze the price variation in cesarean sections, using the de-identified minimum and maximum negotiated prices for the procedure. In this analysis, we began to explore the issue of variables which correlate to price disparities.

**Data Used**

For our analysis, we used a hand-compiled data set of hospital pricing information from 108 hospitals, which we acquired on hospital websites via Google search (Hospital Name + "transparency file"). To our knowledge, no compiled hospital transparency pricing data existed prior to our study. This was a significant challenge, as in many cases hospitals had not yet posted their price transparency information. To supplement our data with income and wealth information, we pulled data from the Census Bureau, aggregated at the zip code level. Using data on rurality provided by the University of Michigan, we classified hospitals as being either rural or urban based upon their connection to urban areas.[16] All data was aggregated into a single file through an inner join on zip code.

We used Diagnostic Related Group (DRG) codes to specify comparable procedures across hospitals. When a patient is admitted to the hospital, most of the time the facility is reimbursed based on the DRG, which accounts for the patient's condition and procedures performed. Out of the six C-section DRGs, reflecting varying combinations of complications/comorbidities and the presence of sterilization (such as tubal ligation), we chose to focus on DRG 788: Cesarean section without sterilization, without complications or comorbidities (CC/MCC). This is because it may be more common among the C-section services. To be included in our study, a hospital had to have transparency data reported with this DRG code included. Many hospitals' transparency files were rejected from analysis because they did not have this DRG code in their data or were reporting the data only by Current Procedural Terminology (CPT), which is a professional reimbursement code and would not include all costs.

---

[15] *Hospital Price Transparency Frequently Asked Questions (FAQS)*, CMS, January 15, 2021, <https://www.cms.gov/files/document/hospital-price-transparency-frequently-asked-questions.pdf>. Last accessed April 25, 2022.

[16] University of Michigan, Popular Studies Center: Institute for Social Research, 2020, <https://www.psc.isr.umich.edu/dis/data/kb/answer/1102.html>

**Methodology**

The approach to data gathering for this study was to cover as large of a geographic area as possible instead of focusing on a specific region of the United States in order to get a representative sample. As stated above, finding hospitals that are fully compliant with the CMS requirement has proven to be a challenge, especially since no list of compliant hospitals has been published to our knowledge. Expanding the geographic area has allowed us to increase the sample size. This means due to the nascent nature of this data we had to search hospital-by-hospital looking for usable data to generate our own dataset for analysis. After excluding some observations due to missing zip code level data, our data set contains 119 hospitals in 26 states.

Figure 2 below shows for which states we have data in our dataset, with a blue circle in a state indicating that we have usable data from that state. The size of the circle represents how much data in our final set is from that state relative to other states. For example, we have data from 16 hospitals from the state of California in our data, which makes up 13.45% of our usable observations, and therefore is a relatively large circle.

**Figure 2: Number of Data Observations By State**

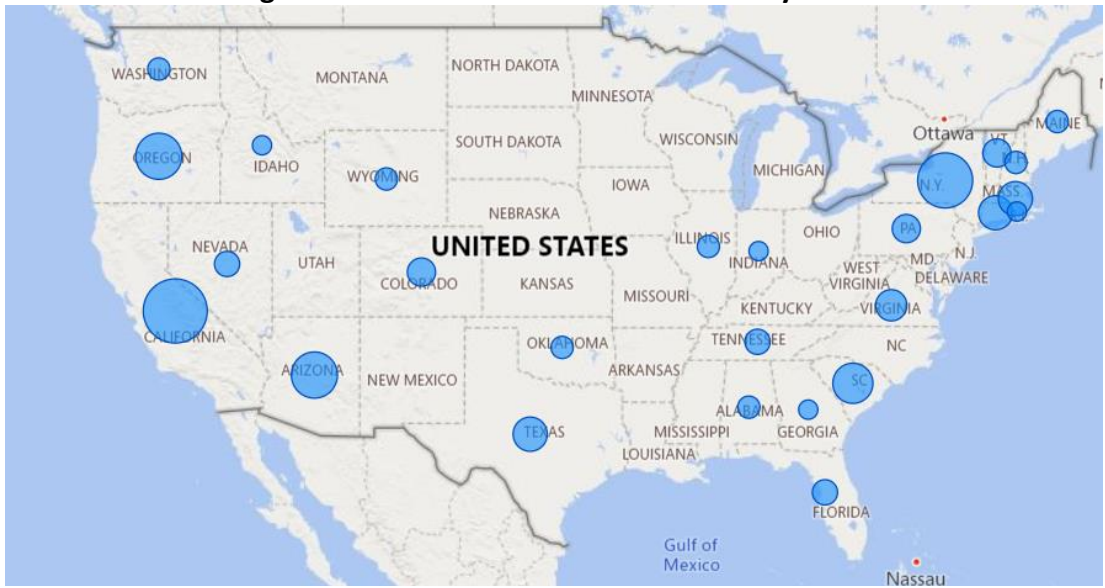

After performing a thorough exploratory data analysis of variables and outlier detection/removal, we used linear regression analysis. Outlier detection was especially important for our relatively small sample size, as leaving outliers in can affect the quality of estimates as well as the overall significance. Linear regression is useful when modeling the change in a dependent "y" variable (in this case, hospital negotiated price ranges) associated with the change in one or more independent "x" (explanatory) variables. Though regression is often used to attempt to establish a causal linkage between two variables, for the purposes of our analysis an association was satisfactory.

**Dependent variable: Negotiated Price Range**
Ideally, to create the dependent variable that adequately captures price variability, we would use pricing data for each individual C-section case in our sampled hospitals within a certain timeframe, which would allow us to calculate the variance around the mean price for the DRG. Such data is not available under the CMS pricing transparency requirements; however, we have the ability to construct a simpler measure of variability, based on the minimum and maximum prices the hospital has negotiated with the third-party payers, i.e. insurance companies and CMS. We are calculating the negotiated price range as the difference between the maximum and minimum negotiated prices for DRG 788. Figure 3 below shows the distribution of the dependent variable: across all sampled hospitals, the average of the negotiated price range was $16,399, with the minimum of $93 and the maximum of $108,130.

**Figure 3: Histogram of the Differences in the Maximum and Minimum Negotiated Prices for DRG 788, at Hospital Level**

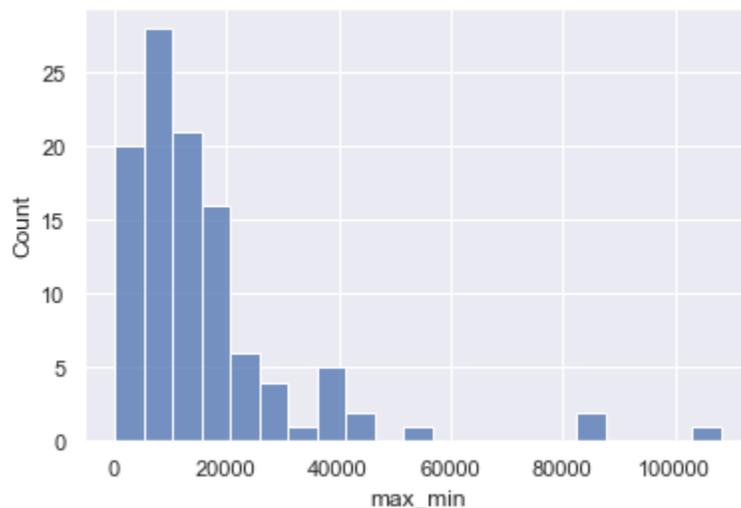

In addition to the difference between the negotiated maximum and minimum prices for cesarean sections, we also explored the use of the quotient of the maximum divided by the minimum negotiated price. Both of these models were near-comparable when comparing levels of significance. However, when comparing the two we felt that the negotiated price range was more robust to the key assumptions of linear regression: a subtracted difference is reflected in US dollars, whereas it is harder to precisely say what a ratio value indicates. Ratio values could additionally be subject to outlier values and/or high leverage points, especially in hypothetical cases where the minimum negotiated value is quite low and the maximum value is quite high. Thus, we ultimately chose to focus on the models which used the difference between the maximum and minimum contracted price.

**Explanatory Variables**

To our knowledge, there are no prior studies to examine the determinants of the hospital price variability based on the compiled transparency data. We selected independent variables based on the assumption that factors correlated to price variability are similar to those that correlate to the level of hospital pricing:

1. *Hospital Size*
   For this variable we are using the total number of inpatient beds in a hospital, made available by the American Hospital Directory (AHD). One of the reasons hospital size may be affecting the pricing is the negotiating power: larger hospitals, attracting a lot of patients, may be able to negotiate higher rates with the insurance companies. To the degree that these higher rates are closer to the undiscounted gross changes, the price variability may be lower. The other reason is economies of scale: by lowering the administrative costs, a hospital is able to charge lower prices. However, if the market dynamics such as supply vs. demand relationship in a certain area allow prices to be higher, that may result in the increased variability, depending on the insurance company's negotiating strength. The bed size of hospitals in our data ranged from 19 to 1639, with a median of 188. It appears that smaller hospitals are more likely to publish the transparency files, possibly because the penalty for non-compliance is more impactful for them than for the large hospitals.

2. *Hospital Quality*
   Although there are many possible ways to measure hospital quality, such as readmissions, patient safety indicators, patient reviews, etc., in our study we have decided to use a binary variable identifying hospitals listed among the Top 20 U.S. News Honor Roll hospitals. Every year U.S. News and World report publishes a list of the top performing hospitals across multiple specialties, based on patient outcomes and other data. Since there is no maternity related specialty, we are using the "Best Hospitals Honor Roll," which takes into account the full range of care and typically includes hospitals that rank among the best across multiple specialties. In a slight deviation from the random hospital sampling, we made an effort to collect the transparency data from as many Honor Roll hospitals as possible, resulting in ten top quality hospitals being included in our data set.

   We hypothesized that quality is likely to affect hospital prices due to (1) higher use of resources in higher quality care and (2) the most reputable hospitals are better able to negotiate higher prices with insurance carriers. This may result in an increase of the maximum negotiated price, which could increase the spread between the maximum and minimum prices. (Alternatively, high quality hospitals may be able to raise their minimum negotiated prices closer to the maximum, causing the price variability to actually diminish.)

3. *Teaching Status*
   Hospital teaching status may affect pricing variability in a similar way as hospital quality. There are additional resources incurred by a hospital to have medical students attend procedures and training. At the same time, it also gives a hospital a good reputation of providing high quality care. The correlation between teaching status and quality of care has been confirmed by some studies, as summarized by J. Ayanyan and JS Weissman.[17] Teaching status data is available through hospital profiles published by the American Hospital Directory. From the 108 sampled hospitals included in final results, 50 are teaching hospitals. We used an indicator variable to classify these variables in our analysis, with teaching hospitals being assigned a value of "1' and non-teaching hospitals being assigned a value of "0."

4. *Urban vs. Rural*
   Using the hospital zip code, we are bringing in the data to indicate whether a hospital is located in an Urban or Rural area. Rurality of the area may impact hospital costs and variability through the economy of scale factor (urban area hospitals would have lower costs thanks to the economies of scale) or due to case mix difference, as urban hospitals tend to attract patients with higher severity and complexity.[18] 28 hospitals in our data sample were flagged as located in a rural area. We used an indicator variable to classify these variables in our analysis, with rural hospitals being assigned a value of "1' and urban hospitals being assigned a value of "0."

5. *Area Gini Coefficient*
   The Gini coefficient is an indicator of income inequality in the area, with the possible range from 0 (low inequality) to 1 (high inequality). By including this variable, we hypothesized that areas with higher income inequality are likely to observe higher price variability, as different income groups are willing to pay for services at different levels of pricing. Gini coefficient for zip codes of the hospitals in our data sample ranged from

---

[17] Ayanian, J. Z., & Weissman, J. S. (2002). Teaching hospitals and quality of care: a review of the literature. *The Milbank quarterly*, *80*(3), 569–v. <https://doi.org/10.1111/1468-0009.00023>.

[18] Street, Andrew & Scheller-Kreinsen, David & Geissler, Alexander & Busse, Reinhard. (2010). Determinants of hospital costs and performance variation. <https://www.researchgate.net/publication/308965668_Determinants_of_hospital_costs_and_performance_variation>

0.36 to 0.66, with a median of 0.45.

6. *Area Median Income*

    Median income of the hospital surrounding area may directly correlate with the level of hospital pricing, as both goods and services tend to cost more in areas with higher income levels. This variable is likely to capture the geographic cost differences, to some degree controlling for hospital resource costs. The values in our data range from $23,500 to $156,000, with a median of $61,318.

**Results**

After performing our exploratory analysis, we began the process of running our regression models. To get to our final model, we first ran a regression model that included all of the dependent variables which we describe above, i.e., quality, rural/urban, gini coefficient, teaching, number of beds, and median income. The results of that model showed us that only the coefficients for quality and teaching were statistically significant. The output for that model can be seen below in Table 1.

**Table 1: Regression Results using all Explanatory variables**

```
========================================================================
                  coef       std err         t      P>|t|     [0.025      0.975]
------------------------------------------------------------------------
Intercept      1.672e+04    1.55e+04       1.081    0.282    -1.4e+04    4.74e+04
rural[T.1]    -6729.4535    4157.533      -1.619    0.109    -1.5e+04    1517.975
Quality        1.837e+04    7547.421       2.435    0.017     3402.935   3.33e+04
Beds             -3.8674       6.754      -0.573    0.568      -17.266      9.532
Teach          7406.8631    3497.396       2.118    0.037      468.970   1.43e+04
median_income     0.0012       0.060       0.020    0.984       -0.117      0.119
gini          -5336.2255    2.96e+04      -0.180    0.857    -6.41e+04    5.35e+04
========================================================================
```

We then iteratively removed explanatory variables that were not statistically significant, evaluating at a p-value = 0.1. We chose this relatively high p-value coefficient (as opposed to 0.05 and 0.01) to accommodate our smaller sample size, which could hamper our ability to find significance at a lower p-value. Our final results demonstrated statistical correlation between a hospital's rurality, its quality, and whether or not it was a teaching hospital. The results can be seen below in Table 2.

**Table 2: Regression Results using the Statistically Significant Explanatory variables**

```
                            OLS Regression Results
==============================================================================
Dep. Variable:                max_min   R-squared:                       0.165
Model:                            OLS   Adj. R-squared:                  0.149
Method:                 Least Squares   F-statistic:                     10.27
Date:                Sat, 14 May 2022   Prob (F-statistic):           8.52e-05
Time:                        13:15:35   Log-Likelihood:                -1182.8
No. Observations:                 107   AIC:                             2372.
Df Residuals:                     104   BIC:                             2380.
Df Model:                           2
Covariance Type:            nonrobust
==============================================================================
                 coef    std err          t      P>|t|      [0.025      0.975]
------------------------------------------------------------------------------
Intercept      1.1e+04   2054.455      5.352      0.000    6920.953    1.51e+04
Quality       1.69e+04   6321.727      2.674      0.009    4365.424    2.94e+04
Teach         8630.8787  3133.014      2.755      0.007    2417.995    1.48e+04
==============================================================================
Omnibus:                       70.541   Durbin-Watson:                   1.519
Prob(Omnibus):                  0.000   Jarque-Bera (JB):              350.831
Skew:                           2.247   Prob(JB):                     6.58e-77
Kurtosis:                      10.648   Cond. No.                         4.83
==============================================================================
```

Since the coefficients on both Quality and Teaching are positive and statistically significant, this model tells us that as if a hospital is a teaching hospital then the price difference between maximum and minimum price on a C-section procedure is higher than if a hospital is not a teaching hospital, and similarly if a hospital is classified as a US News Honor Roll hospital it has a higher difference than if it is not in that list. More explicitly, a teaching hospital is correlated to increase in the range of about $7,000 (indicating more price variability), and a top hospital is associated with an increase in this range of about $16,000. Our results had a relatively low R-squared value of 0.165, interpreted as hospital quality and whether the hospital was a teaching hospital only explaining about 17% of the variability in the price difference.

**Follow up**

This analysis represents an attempt to analyze hospitals' transparency data to understand price variability based on one DRG code. There is a significant opportunity for the expansion and refinement of this study. The primary issues we note are sample size, sample quality, explanatory variables, and using payer-specific negotiated prices.

1. *Increased Sample Size*
   First and foremost, given more time and resources for data collection, we could significantly expand the sample size, which would improve the credibility of our results. Also, expanding the data to include a variety of services, both by DRG and CPT codes, would make the findings around hospital price variability more generalizable. Further, if the transparency data is collected over multiple years, the resulting panel data would allow analysis of the dynamics of changes in explanatory variables.

2. *Hospital Compliance and Sample Selection Bias*
Hospitals have been slow to comply with the CMS requirement. In a March 2021 analysis, researchers found in a random sampling that 65 percent of hospitals were noncompliant.[19] Hospitals have cited a myriad of reasons for not publishing prices including the overwhelming medical and administrative burden of dealing with the influx of COVID cases during the pandemic, not wanting to provide public pricing information competitors, these data not accurately reflecting prices paid so it isn't useful, and this hospital providing a cost-calculator tool on the site and therefore this is sufficient.

In addition to a flat-out refusal, perhaps as problematic for empirical research purposes as not producing any data was that some hospitals produced data that were unusable for a differing degree of reasons. There were hospitals that did not produce the data in a machine-readable file as specified by CMS but instead presented the data in an interactive dashboard-like environment. Certain hospitals did not produce all the requested fields and others that did not produce data that covered all the necessary items and services. Some hospitals that were in the same hospitals system, meaning that multiple hospitals were part of the same organization with different geographical locations reported the same exact minimum negotiated price and maximum negotiated price – down to a penny – for the same procedure. This could be an accurate representation of pricing, meaning that Hospital A, Hospital B, and Hospital C all charged the same minimum and maximum amount for the same procedure an should be evaluated individually as valid observations, but it also could be the same minimum and maximum price for one procedure reported across all entities of the larger group of hospitals and therefore should not be separately valid transactions; it is impossible to tell from these data.

A later study done in December 2021 estimated that 55 percent of hospitals were not in compliance.[20] This study made a salient observation that a hospital's likelihood to be compliant was strongly correlated with the other hospitals' compliance in the same regional market. This held true for purposes of this study, found this to be anecdotally true when searching for useful data in this study. In addition to compliance correlations, it was also notable that some hospitals that were unrelated by anything other than geography, i.e., they were not part of the same hospital system, but they were geographically similar, had similar reporting structure to their data files. In fact, the December 2021 study goes on to suggest that "hospitals do not make decisions in isolation, rather their decisions reflect market pressure from their peers"[21] which we also found to be accurate during the data collection process for this study, i.e., if we found useful data for a hospital in a given state we were likely to more easily find other hospitals with useful data.

---

[19] Low Compliance From Big Hospitals On CMS's Hospital Price Transparency Rule. (2021). Forefront Group. <https://www.healthaffairs.org/do/10.1377/forefront.20210311.899634>.

[20] Study Estimates That More Than Half of U.S. Hospitals Not in Compliance With New Pricing Disclosure Rules in First Five Months | Johns Hopkins. (2021, December 9). Johns Hopkins Bloomberg School of Public Health. <https://publichealth.jhu.edu/2021/study-estimates-that-more-than-half-of-us-hospitals-not-in-compliance-with-new-pricing-disclosure-rules-in-first-five-months>.

Ultimately, these challenge the conclusions of our model by way of not knowing what we didn't capture. This is what is known as sample selection bias, which occurs when a sampled group meaningfully differs in characteristics that alter the statistical significance of results. If a hospital did not comply for reasons of administrative burden or COVID-related difficulties (which would likely not alter our results significantly), for example, that is different than wilful non-compliance by hospitals that charged higher prices (which would alter our results). In order for statistical analysis to be meaningful, generally the sample must be representative.

3. *Improved Explanatory Variables*
There is also a significant opportunity for explanatory variable improvement. We found that the hospital ranking and teaching status are statistically significant; we could further understand the impact of quality using a continuous variable such as the Total Performance Score or even each of the domains it is based on: clinical outcomes, patient safety, community engagement, and efficiency and cost reduction.[22] Also, with the increase in sample size we could introduce dummy variables for different states or MSAs, which would allow us to understand and control for state specifics such as policy regulations.

4. *Using Payer-Specific Negotiated Prices*
One of the most significant potential improvements to the model is related to making use of the payer-specific negotiated prices, which is one of the CMS transparency requirements. Most of the hospitals in the files we obtained did not report the data by payer, limiting the number to the de-identified minimum and maximum contracted amounts. Chances are that even the number of payers that a hospital is contracted with would explain a lot of the price variability: hospitals contracting with a handful of insurance companies would have a much lower variability than facilities with many payer contracts. Further, it would be valuable to understand if certain insurance companies allow very high or very low negotiated charges, creating higher variability.

---

[21] Ibid

[22] Centers for Medicare & Medicaid Services. (2019, October 29). Fact sheet CMS hospital value-based purchasing program results for Fiscal Year 2020. CMS. Retrieved May 14, 2022, from https://www.cms.gov/newsroom/fact-sheets/cms-hospital-value-based-purchasing-program-results-fiscal-year-2020

**Practical Implications**

The data we have collected confirms drastic variability in pricing for the same procedure: at a hospital level, price difference for the same procedure can differ by over $100,000 depending on which insurance is covering your C-section and what hospital you have chosen for the procedure. On average, there is a $16,399 price difference between the maximum and minimum negotiated prices within a hospital.

While consumers covered by health insurance do not bear the full burden of prices charged by a hospital, this variability is likely to correlate with the higher uncertainty around the out-of-pocket costs, depending on the insurance benefit structure. Therefore, both hospitals and insurance companies need to enhance their efforts to improve pricing transparency, not just pursuant to CMS requirements, but in day-to-day operations, to make sure that a consumer gets as accurate an estimate of the procedure cost as possible.

Consumers that have a strong preference for a top-ranked teaching facility to perform their procedure and have the option to choose from multiple insurance companies should use the hospital cost estimators along with the benefit design information to choose their insurance coverage plan. Consumers that do not have a strong preference around the hospital status may be better off choosing to have a relatively standard procedure performed at a less acclaimed local hospital.

Both hospitals and insurance companies, i.e. the negotiating parties in setting the price per procedure, can use the transparency data to see what their competitors are charging or paying, which will tend to reduce the price variability ranges. In this respect it is very important that the transparency effort continues, correcting the market failure to set a fair price due to imperfect information.

**References**

1. Centers for Medicare & Medicaid Services. (2019, October 29). Fact sheet CMS hospital value-based purchasing program results for Fiscal Year 2020. CMS. Retrieved May 14, 2022, from https://www.cms.gov/newsroom/fact-sheets/cms-hospital-value-based-purchasing-program-results-fiscal-year-2020

2. Low Compliance From Big Hospitals On CMS's Hospital Price Transparency Rule. (2021). Forefront Group. <https://www.healthaffairs.org/do/10.1377/forefront.20210311.899634>.

3. Whieldon, Lee, and Huthaifa I. Ashqar. "Predicting residential property value: a comparison of multiple regression techniques." SN Business & Economics 2.11 (2022): 178.

4. Study Estimates That More Than Half of U.S. Hospitals Not in Compliance With New Pricing Disclosure Rules in First Five Months | Johns Hopkins. (2021, December 9). Johns Hopkins Bloomberg School of Public Health. <https://publichealth.jhu.edu/2021/study-estimates-that-more-than-half-of-us-hospitals-not-in-compliance-with-new-pricing-disclosure-rules-in-first-five-months>.

5. Radwan, Ahmad, et al. "Predictive analytics in mental health leveraging llm embeddings and machine learning models for social media analysis." International Journal of Web Services Research (IJWSR) 21.1 (2024): 1-22.

6.  Ayanian, J. Z., & Weissman, J. S. (2002). Teaching hospitals and quality of care: a review of the literature. The Milbank quarterly, 80(3), 569–v. <https://doi.org/10.1111/1468-0009.00023>.

7.  Street, Andrew & Scheller-Kreinsen, David & Geissler, Alexander & Busse, Reinhard. (2010). Determinants of hospital costs and performance variation. https://www.researchgate.net/publication/308965668_Determinants_of_hospital_costs_and_performance_variation

8.  Emanuel, Ezekiel, et al. "A Systemic Approach to Containing Health Care Spending." New England Journal of Medicine, vol. 367, no. 10, 2012, pp. 949–54. <https://www.nejm.org/doi/full/10.1056/NEJMsb1205901>.

9.  How New Data On Hospital "Discounted Cash Prices" Might Lead To Patient Savings. (2021). Forefront Group. <https://www.healthaffairs.org/do/10.1377/forefront.20211103.716124>.

10. U.S. Census Bureau. (2021, December 16). 19% of U.S. Households Could Not Afford to Pay for Medical Care Right Away. Census.Gov. <https://www.census.gov/library/stories/2021/04/who-had-medical-debt-in-united-states.html>.

11. Kluender R, Mahoney N, Wong F, Yin W. Medical Debt in the US, 2009-2020. JAMA. 2021;326(3):250–256. <https://jamanetwork.com/journals/jama/fullarticle/2782187>.

12. National Expenditures Highlights, CMS, 2020, <https://www.cms.gov/files/document/highlights.pdf>. Last accessed April, 28, 2022.

13. National Health Expenditure Accounts, CMS, 2020. <https://www.cms.gov/Research-Statistics-Data-and-Systems/Statistics-Trends-and-Reports/NationalHealthExpendData/NationalHealthAccountsHistorical>. Last accessed May, 9, 2022.

14. Woodard, Davon, Huthaifa I. Ashqar, and Taoran Ji. "Ethics, Data Science, and Health and Human Services: Embedded Bias in Policy Approaches to Teen Pregnancy Prevention." arXiv preprint arXiv:2006.04029 (2020).